# Orbit optimization for ASTROD-GW and its time delay interferometry with two arms using CGC ephemeris

G. Wang[a),b)] and W.-T. Ni[c),d)]

[a)] *Shenzhen National Climate Observatory, No.1 Qixiang Rd., Zhuzilin, Futian District, Shenzhen, 518040, China*

[b)] *Purple Mountain Observatory, Chinese Academy of Sciences, No. 2 Beijing W. Rd., Nanjing, 210008, China*

[c)] *Shanghai United Center for Astrophysics (SUCA), Shanghai Normal University, 100 Guilin Road, Shanghai, 200234, China*

[d)] *Center for Gravitation and Cosmology (CGC), Department of Physics, National Tsing Hua University, Hsinchu, Taiwan, 300, ROC*

e-mails: gwanggw@gmail.com, weitou@gmail.com

**Abstract**

ASTROD-GW (ASTROD [Astrodynamical Space Test of Relativity using Optical Devices] optimized for Gravitation Wave detection) is an optimization of ASTROD to focus on the goal of detection of gravitation waves. The detection sensitivity is shifted 52 times toward larger wavelength compared to that of LISA. The mission orbits of the 3 spacecraft forming a nearly equilateral triangular array are chosen to be near the Sun-Earth Lagrange points L3, L4 and L5. The 3 spacecraft range interferometrically with one another with arm length about 260 million kilometers. In order to attain the requisite sensitivity for ASTROD-GW, laser frequency noise must be suppressed below the secondary noises such as the optical path noise, acceleration noise etc. For suppressing laser frequency noise, we need to use time delay interferometry (TDI) to match the two different optical paths (times of travel). Since planets and other solar-system bodies perturb the orbits of ASTROD-GW spacecraft and affect the (TDI), we simulate the time delay numerically using CGC 2.7 ephemeris framework. To conform to the ASTROD-GW planning, we work out a set of 20-year optimized mission orbits of ASTROD-GW spacecraft starting at June 21, 2028, and calculate the residual optical path differences in the first and second generation TDI for one-detector case. In our optimized mission orbits for 20 years, changes of arm length are less than 0.0003 AU; the relative Doppler velocities are less than 3m/s. All the second generation TDI for one-detector case satisfies the ASTROD-GW requirement.

Submitted May 23, 2012

## 1. Introduction

The gravitation-wave (GW) experimental groups in the world are actively looking forward to GW detection in the near future.[1] The working detectors at present are mainly ground interferometers and Pulsar Timing Arrays (PTAs). The first-generation ground interferometers include LIGO (4 km arm length),[2] Virgo (3 km arm length),[3] GEO (0.6 km arm length)[4] and TAMA (0.3 km arm length).[5] The second-generation ground interferometers under construction are Advanced LIGO,[6] Advanced Virgo[7] and KAGRA/LCGT (3 km arm length).[8] LIGO-India is under active consideration.[9] KAGRA is an underground detector and its main mirrors will become cryogenic in its second phase to suppress thermal noises. When the second- generation interferometers are completed and their sensitivity goals in the high frequency band (10 Hz – 100 kHz; for a classification of GWs, see references [1, 10]) are achieved in about 5 years, first direct detection of GWs from binary neutron-star mergers, neutron-star/black-hole mergers or black-hole/black-hole mergers will be realized. PTAs seek for detection of GWs from supermassive black hole merger events and background in the very low frequency band (300 pHz – 100 nHz) around 2020.[11]

To detect GWs from various different astrophysical and cosmological sources in different spectral ranges and to enhance the signal to noise ratios, we need also to explore the GW spectrum between the high frequency band and the very low frequency band, i.e., the middle frequency band (0.1 Hz – 10 Hz) and the low frequency band (100 nHz – 0.1 Hz). Space detectors are most sensitive to these bands. Mission concepts under implementation and study are LISA,[12] ASTROD,[13,14] Super-ASTROD,[15] ASTROD-GW,[16,17] BBO[18] and DECIGO.[19,20] Except for the Fabry-Perot implementation of DECIGO whose scheme is basically like the ground GW interferometric detectors, all other missions have unequal arm lengths and laser frequency noise is very serious. One way to suppress it is to use time delay interferometry (TDI) by combining paths to make the two interfering beam to have closely equal optical paths. Time delay interferometry is considered for ASTROD in 1996[21,22] and has been worked out for LISA much more thoroughly since 1999.[23,24]

With the NASA withdrawal in April 2011, the LISA project was nominally ended. A joint effort from seven European countries (France, Germany, Italy, The Netherlands, Spain, Switzerland, UK) and ESA have made a new proposal NGO/eLISA by down-scaling the size and the arm length of LISA. The mission orbit configuration of S/Cs is similar to LISA with the configuration plane inclined by about 60° to the ecliptic, but with nominal arm length $1 \times 10^6$ km (instead of $5 \times 10^6$ km) and the configuration trailing Earth by 10-20° (instead of around 20°). The mission duration is for 2 years mission (science) orbit (about 4 years including transferring and commissioning). NGO/eLISA is one detector with two arms.[25]

Dhurandhar, Nayak and Vinet[26] have worked out the time delay interferometry observables for LISA for which only the data streams from two arms are considered. These TDI observables could be readily applied to the similar situation for ASTROD-GW (nominal arm length 260 × $10^6$ km) and to the case of NGO/eLISA which is already a two-arm interferometer. Following the semi-analytical work of Dhurandhar, Nayak and Vinet,[26] we have worked out these TDI observables in a series of *numerical studies of TDIs* for LISA (nominal arm length 5 × $10^6$ km),[27] NGO/eLISA[28] and for an NGO-LISA-type mission with a nominal arm length of 2 × $10^6$ km.[28]

In this paper, we work out the time delay interferometry for ASTROD-GW in the case of one-detector two-arm case. In section 2, we review the ASTROD-GW mission concept. In section 3, we summarize CGC 2.7 ephemeris framework to be used for optimizing the mission orbit design and to apply to the second-generation TDIs. In section 4, we optimize the mission orbit of ASTROD-GW to have nearly equal arm lengths and to have minimal line-of-site Doppler velocity between different pairs of spacecraft for 20 years. In section 5, we summarize the basics of time-delay interferometry. In section 6, we work out the first generation time delay interferometry for ASTROD-GW numerically. In section 7, we work out the second generation TDI for ASTROD-GW with one interferometer. In section 8, we conclude this paper with discussions. Some of the results of this paper have been reported earlier in an unpublished report[29] and in a master thesis.[30] Part of the results are also used in the comparison of TDIs for NGO/eLISA, for an NGO-LISA-type mission with a nominal arm length of 2 × $10^6$ km, for LISA and for ASTROD-GW.[28]

## 2. ASTROD-GW

ASTROD-GW (ASTROD [Astrodynamical Space Test of Relativity using Optical Devices] optimized for GW detection) is an optimization of ASTROD to focus on the goal of detection of GWs.[1,10,16,17,31] The scientific aim is focused on GW detection at low frequency. The mission orbits of the 3 spacecraft forming a nearly equilateral triangular array are chosen to be near the Sun-Earth Lagrange points L3, L4 and L5 (Figure 1).

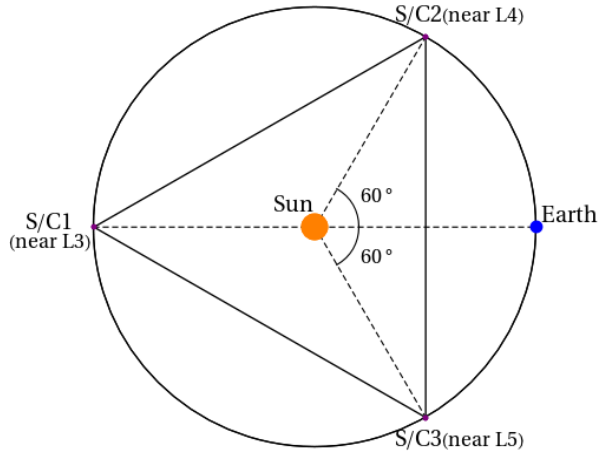

Fig. 1. Schematic of ASTROD-GW mission orbit design.

The 3 spacecraft range interferometrically with one another with arm length about 260 million kilometers. After a first mission-orbit optimization[32,33], the changes of arm length are less than 0.0003 AU or, fractionally, less than ± 1.7 × 10$^{-4}$ in ten years, and the Doppler velocities for the three spacecraft are less than ±4 m/s. These parameters are consistent with those of LISA requirement and, therefore, a number of technologies developed by LISA could be applied to ASTROD-GW also. For the purpose of primordial GW detection, a 6-S/C formation for ASTROD-GW will be used for correlated detection of stochastic GWs.[1,10,16,17,31]

Since the arm length is longer than LISA by 52 times, with 1-2 W laser power and LISA acceleration noise, the strain sensitivity of ASTROD-GW is 52 times lower than LISA, and is better than LISA and Pulsar Timing Arrays (PTAs) in the frequency band 100 nHz - 1 mHz. The sensitivity curve for ASTROD-GW as compared with LISA is shown in Figure 2.[10,31] With longer arm length, the requirement for the power of weak light phase locking is more stringent for ASTROD-GW. In the case of LISA and NGO/eLISA, they are around 100 pW and 1 nW respectively; in the case of ASTROD-GW, it is in the range of 50-100 fW. We have demonstrated the weak light phase locking capability to 2 pW.[34,35] More recently, JPL people have demonstrated the weak light phase locking capability to 40 f W.[36] The power requirement feasibility of phase locking is met. ASTROD-GW will complement NGO/eLISA and PTA in exploring black hole co-evolution with galaxies and GW backgrounds in the important frequency range 100 nHz - 1 mHz.

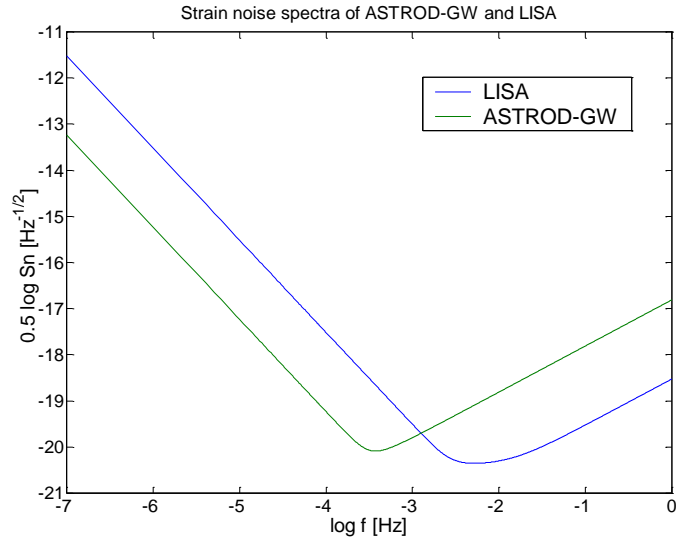

Fig. 2. ASTROD-GW strain noise amplitude spectrum as compared to LISA strain noise amplitude spectrum.

## 3. CGC ephemeris

In 1998, we started orbit simulation and parameter determination for ASTROD.[37,38] We worked out a post-Newtonian ephemeris of the Sun including the solar quadrupole moment, the major planets and 3 biggest asteroids. We term this working ephemeris CGC 1 (CGC: Center for Gravitation and Cosmology). Using this ephemeris as a deterministic model and adding stochastic terms to simulate noise, we generate simulated ranging data and use

Kalman filtering to determine the accuracies of fitted relativistic and solar-system parameters for 1050 days of the ASTROD mission.

For a better evaluation of the accuracy of Ġ/G, we need also to monitor the masses of other asteroids. For this, we considered all known 492 asteroids with diameter greater than 65 km to obtain an improved ephemeris framework --- CGC 2, and calculated the perturbations due to these 492 asteroids on the ASTROD spacecraft.[39,40]

In building CGC ephemeris framework, we use the post-Newtonian barycentric metric and equations of motion as derived in Brumberg[41] with PPN (Parametrized Post-Newtonian) parameters β and γ for solar system bodies. The metric with the gauge parameter α set to zero is

$$ds^2 = [1 - 2\sum_i \frac{m_i}{r_i} + 2\beta(\sum_i \frac{m_i}{r_i})^2 + (4\beta - 2)\sum_i \frac{m_i}{r_i}\sum_{j\neq i}\frac{m_j}{r_{ij}}$$
$$- c^{-2}\sum_i \frac{m_i}{r_i}(2(\gamma+1)\dot{x}_i^2 - r_i \cdot \ddot{x}_i - \frac{1}{r_i^2}(r_i \cdot \dot{x}_i)^2) + \frac{m_1 R_1^2}{r_1^3} J_2(3(\frac{r_1 \cdot \hat{z}}{r_1})^2 - 1)]c^2 dt^2$$
$$+ 2c^{-1}\sum_i \frac{m_i}{r_i}((2\gamma+2)\dot{x}_i) \cdot d\boldsymbol{x} c dt - [1 + 2\gamma\sum_i \frac{m_i}{r_i}](d\boldsymbol{x})^2 \quad (1)$$

where $r_i = \boldsymbol{x} - \boldsymbol{x}_i$, $r_{ij} = \boldsymbol{x}_i - \boldsymbol{x}_j$, $m_i = GM_i/c^2$, and $M_i$'s the masses of the bodies with $M_1$ the solar mass.[41] $J_2$ is the quadrupole moment parameter of the Sun. $\hat{z}$ is the unit vector normal to the elliptic plane. The associated equations of motion of N-mass problem as derived from the geodesic variational principle of this metric are

$$\ddot{\boldsymbol{x}}_i = -\sum_{j\neq i} \frac{GM_j}{r_{ij}^3}\boldsymbol{r}_{ij} + \sum_{j\neq i} m_j(A_{ij}\boldsymbol{r}_{ij} + B_{ij}\dot{\boldsymbol{r}}_{ij})$$
$$A_{ij} = \frac{\dot{x}_i^2}{r_{ij}^3} - (\gamma+1)\frac{\dot{r}_{ij}^2}{r_{ij}^3} + \frac{3}{2r_{ij}^5}(r_{ij}\dot{x}_j)^2 + G[(2\gamma+2\beta+1)M_i + (2\gamma+2\beta)M_j]\frac{1}{r_{ij}^4} \quad (2)$$
$$+ \sum_{k\neq i,j} GM_k[(2\gamma+2\beta)\frac{1}{r_{ij}^3 r_{ik}} + (2\beta-1)\frac{1}{r_{ij}^3 r_{jk}} + \frac{2(\gamma+1)}{r_{ij}r_{jk}^3} - (2\gamma+\frac{3}{2})\frac{1}{r_{ik}r_{jk}^3} - \frac{1}{2r_{jk}^3}\frac{r_{ij}r_{ik}}{r_{ij}^3}]$$
$$B_{ij} = \frac{1}{r_{ij}^3}[(2\gamma+2)(r_{ij}\dot{r}_{ij}) + (r_{ij}\dot{x}_j)]$$

These equations are used to build our computer-integrated ephemeris (with $\gamma = \beta = 1$, $J_2 = 2 \times 10^{-7}$) for eight-planets, the Pluto, the Moon and the Sun. The positions and velocities at the epoch 2005.6.10 0:00 are taken from the DE403 ephemeris. The evolution is solved by using the 4$^{th}$-order Runge-Kutta method with the step size h =0.01 day. In [37] the 11-body evolution is extended to 14-body to include the 3 big asteroids — Ceres, Pallas and Vesta (CGC 1 ephemeris). Since the tilt of the axis of the solar quadrupole moment to the perpendicular of the elliptical plane is small (7°), in CGC 1 ephemeris, we have neglected this tilt. In CGC 2 ephemeris, we have added the perturbations of additional 489 asteroids.

In our previous optimization of ASTROD-GW orbits,[32,33] we have used CGC 2.5 ephemeris in which only 3 biggest minor planets are taken into accounts, but the Earth's precession and nutation are added; the solar quadratic zonal harmonic and the Earth's quadratic to quartic zonal harmonic are considered.

In this paper, we add the perturbation of additional 349 asteroids and call it CGC 2.7 ephemeris.

The differences in orbit evolution compared with DE405 for Earth for 3700 days starting at JD2461944.0 (2028-Jun-21 12:00:00) are shown in Fig. 3. The differences in radial distances are less than about 200 m.

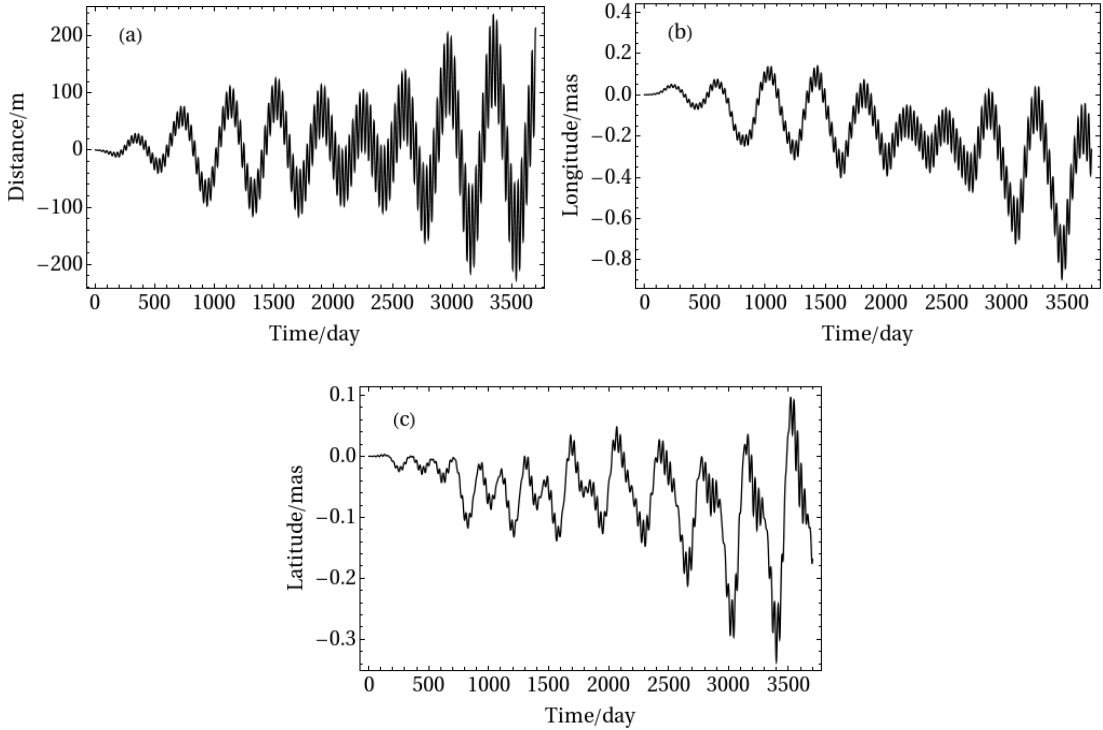

Fig. 3. Differences (DE405-CGC2.7) in (a) radial distance, (b) longitude, (c) latitude starting at JD2461944.0 for Earth orbits in DE405 and CGC2.7 ephemerides in the J2000 Heliocentric Earth mean equatorial coordinate system

## 4. Mission orbit optimization

The goal of ASTROD-GW mission orbit optimization is to equalize the three arm lengths of ASTROD-GW formation and to reduce the relative line-of-sight velocities between three pairs of spacecraft as much as possible. In our previous optimization, the time of start of the science part of the mission is chosen to be June 21, 2025 (JD2460848.0) and the optimization is for a period of 3700 days.[32, 33] Since the preparation of the mission may take longer time and there is a potential that the extended mission life time may be longer than 10 years, in this paper, we start at noon, June 21, 2028 (JD2461944.0) and optimize for a period of 20 years following our previous method.

4.1 Initial choice of spacecraft initial conditions

A possible launch time for ASTROD-GW is in the later part of 2020's. For the sake of definiteness we choose noon, June 21, 2028 (JD2461944.0) as the initial epoch for the mission orbit. Other choice of the initial epoch also has similar solutions. Since the dominant force on the spacecraft is from the Sun in the restricted three-body problem of Earth-Sun-spacecraft system. To start the optimization, the initial positions of the three spacecraft can be chosen to form an equilateral triangle on 1 AU solar orbit and the initial velocity chosen to follow an orbit period of one sidereal year. To fix the orbit phase of the three spacecraft, we notice that L4 and L5 Lagrange points of Sun-Earth system is stable and L3 Lagrange point is mildly unstable with time constant about 50 years. For a mission of 10-20 years, L3 point is actually quasi-stable. Therefore, we choose the initial positions of three spacecraft in the heliocentric ecliptic coordinate system to be

S/C1 at (0, 1 AU, 0), S/C2 at ($3^{1/2}2^{-1}$ AU, -0.5 AU, 0), S/C3 at ($-3^{1/2}2^{-1}$ AU, -0.5 AU, 0).

S/C1 is near L3 point, S/C2 near L4 point, S/C3 near L5 point. Since the Earth-Sun orbit is elliptical, the Lagrange points are not stationary. However, the spacecraft are in the halo orbit of the respective Lagrange points largely compensating the non-stationary motion of the Lagrange points to remain nearly circular orbits of the Sun. The initial velocities of the three spacecraft in the heliocentric ecliptic coordinate system are chosen to be

S/C1 at ($-v_0$, 0, 0), S/C2 at (0.5 $v_0$, $3^{1/2}2^{-1}$ $v_0$, 0), S/C3 at (0.5 $v_0$, $-3^{1/2}2^{-1}$ $v_0$, 0),

where $v_0$ (= 0.01720209895 AU/day) is the circular velocity of spacecraft at 1 AU from the Sun.

The initial positions and initial velocities of the three spacecraft in J2000 equatorial solar-system-barycentric coordinate system are listed in the third column of Table 1 which also lists the initial conditions for an intermediate step (after period optimization) and for final results in our optimization for easy comparison.

4.2 Method of optimization

In the solar system the ASTROD-GW spacecraft orbits are perturbed by the other planets. The largest are due to Jupiter and Venus. Therefore the orbital period and the eccentricity change. Our method of optimization is to modify the initial velocities and initial heliocentric distances so that the perturbed orbital periods for ten or twenty year average remain close to sidereal year and the average eccentricity remain near zero.

Table 1. Initial states of S/Cs at epoch JD2461944.0 for our initial choice, after period optimization, and after all optimizations in J2000 equatorial solar-system-barycentric coordinate system

| | | Initial choice of S/C initial states | Initial states of S/Cs after period optimization | Initial states of S/Cs after final optimization |
|---|---|---|---|---|
| S/C1 Position (AU) | X | 1.15400625657242E-3 | 1.15400625657242E-3 | 1.15400625657242E-3 |
| | Y | 9.15264788418969E-1 | 9.15261701184544E-1 | 9.15289225648841E-1 |
| | Z | 3.96855707171320E-1 | 3.96854368692135E-1 | 3.96866302001196E-1 |
| S/C1 Velocity (AU/day) | Vx | -1.72009645137827E-2 | -1.72008300146903E-2 | -1.72003163872199E-2 |
| | Vy | 4.88112077380618E-6 | 4.88112077380618E-6 | 4.88112077380618E-6 |
| | Vz | 2.07014410548162E-6 | 2.07014410548162E-6 | 2.07014410548162E-6 |
| S/C2 Position (AU) | X | 8.67179410041011E-1 | 8.67179419751799E-1 | 8.67153438989685E-1 |
| | Y | -4.60958426468166E-1 | -4.60958432557285E-1 | -4.60944670325136E-1 |
| | Z | -1.99809745830377E-1 | -1.99809748470332E-1 | -1.99803781815802E-1 |
| S/C2 Velocity (AU/day) | Vx | 8.60218391121732E-3 | 8.60233950755836E-3 | 8.60259754371050E-3 |
| | Vy | 1.36730297780320E-2 | 1.36730847422802E-2 | 1.36734947883888E-2 |
| | Vz | 5.92793451120697E-3 | 5.92795834111164E-3 | 5.92813611775755E-3 |
| S/C3 Position (AU) | X | -8.64871397527866E-1 | -8.64871407921666E-1 | -8.64862747667628E-1 |
| | Y | -4.60958426468166E-1 | -4.60958431471891E-1 | -4.60953844061175E-1 |
| | Z | -1.99809745830377E-1 | -1.99809747999756E-1 | -1.99807759114913E-1 |
| S/C3 Velocity (AU/day) | Vx | 8.60218391121732E-3 | 8.60233303252539E-3 | 8.60242121981651E-3 |
| | Vy | -1.36632675364844E-2 | -1.36633074545593E-2 | -1.36634452887228E-2 |
| | Vz | -5.92379422299601E-3 | -5.92381152958984E-3 | -5.92387128797985E-3 |

The total orbital energy (kinetic and potential) of a planet around the Sun for two-body problem is given by

$$E = \frac{mv^2}{2} - \frac{G(M+m)m}{r} = -\frac{G(M+m)m}{2a} \tag{3}$$

The sidereal orbital period of planet is

$$T = \frac{2\pi a^{3/2}}{\sqrt{G(M+m)}} \tag{4}$$

Here $G$ is the gravitational constant, $M$ is the mass of Sun, $m$ is the planetary mass, $v$ is the orbital velocity of the planet, $r$ the planetary heliocentric distance and $a$ is the semi-major axis of planetary orbit.

4.2.1 Trimming the period

For ASTROD-GW spacecraft, the orbit is near circular and we have $r \approx a$. From (3), we have:

$$v^2 \approx \frac{G(M+m)}{r} \approx \frac{G(M+m)}{a} \tag{5}$$

Two useful expressions between orbital period and velocity, and orbital period and position can be found from the equation (4) and (5):

$$dv = -\frac{1}{3}\frac{dT}{T}v$$
$$dr = \frac{2}{3}\frac{dT}{T}r \tag{6}$$

Actually, we use the equation (6) as

$$\mathbf{V}_{new} = \mathbf{V}_{prev} + \delta\mathbf{V} \approx (1-\frac{1}{3}\frac{dT}{T})\mathbf{V}_{prev}$$
$$\mathbf{R}_{new} = \mathbf{R}_{prev} + \delta\mathbf{R} \approx (1+\frac{2}{3}\frac{dT}{T})\mathbf{R}_{prev} \tag{7}$$

where $\mathbf{V}_{prev}, \mathbf{R}_{prev}$ are the velocity and position of spacecraft before adjusting, $dT$ is the variation of the orbital period wanted. $\mathbf{V}_{new}, \mathbf{R}_{new}$ are the new initial conditions with new period.

4.2.2 Trimming the eccentricity

From the equation (4), we notice that the orbital period $T$ is determined by the semi-major axis $a$. Keeping the period means fixing $a$. From equation (3), the method to adjust the eccentricity keeping the period fixed is to have:

$$\frac{dv}{v} = -\frac{dr}{r} \tag{8}$$

To trim the eccentricity we need to change the initial velocity and initial radial distance using

$$\mathbf{V}_{new} = \mathbf{V}_{prev} + \delta\mathbf{V} \approx (1-\frac{dR}{R})\mathbf{V}_{prev}$$
$$\mathbf{R}_{new} = \mathbf{R}_{prev} + \delta\mathbf{R} \approx (1+\frac{dR}{R})\mathbf{R}_{prev} \tag{9}$$

Here $R$ is the initial heliocentric distance of spacecraft. And the change $dR$ is to induce the variation of eccentricity wanted.

4.3 Optimization steps

At the beginning of an optimization step, we calculate the orbits of the spacecraft from initial conditions using the ephemeris CGC 2.7. From these orbits, we calculate the mean orbital period in 20 years, the variations of arm lengths, differences of arm lengths, heliocentric distances of spacecraft and Doppler velocities between spacecraft as functions of the mission time.

With these data, we use equation (7) to optimize the orbital period of each spacecraft. (i) We first calculate the average periods of each spacecraft and change the initial conditions to match the periods to one sidereal year to minimize the change of arm lengths. (ii) Due to planetary perturbations, for the spacecraft near L3 point, the period decreases slightly as time goes on, while for the spacecraft near L4 and L5 points, the period increases slightly as time goes on; Arm3 changes from short to longer, Arm2 changes from getting longer to getting shorter, and the Arm1 changes are relatively small. In this situation, we decrease the initial velocity of S/C1 to let the initial period slightly longer; and we increase the initial velocities of S/C2 and S/C3 to let their initial period slightly shorter. This way, we can compensate the period deviation tendency of the three spacecraft, and minimize the period differences and arm length differences dynamically. After trimming the period, the arm length differences are decreased; however, in general, the orbit configuration still does not satisfy mission requirements.

In the next step, we use equation (9) to trim the S/C eccentricities to be nearly circular, until initial heliocentric distance getting close to perihelion or aphelion distance.

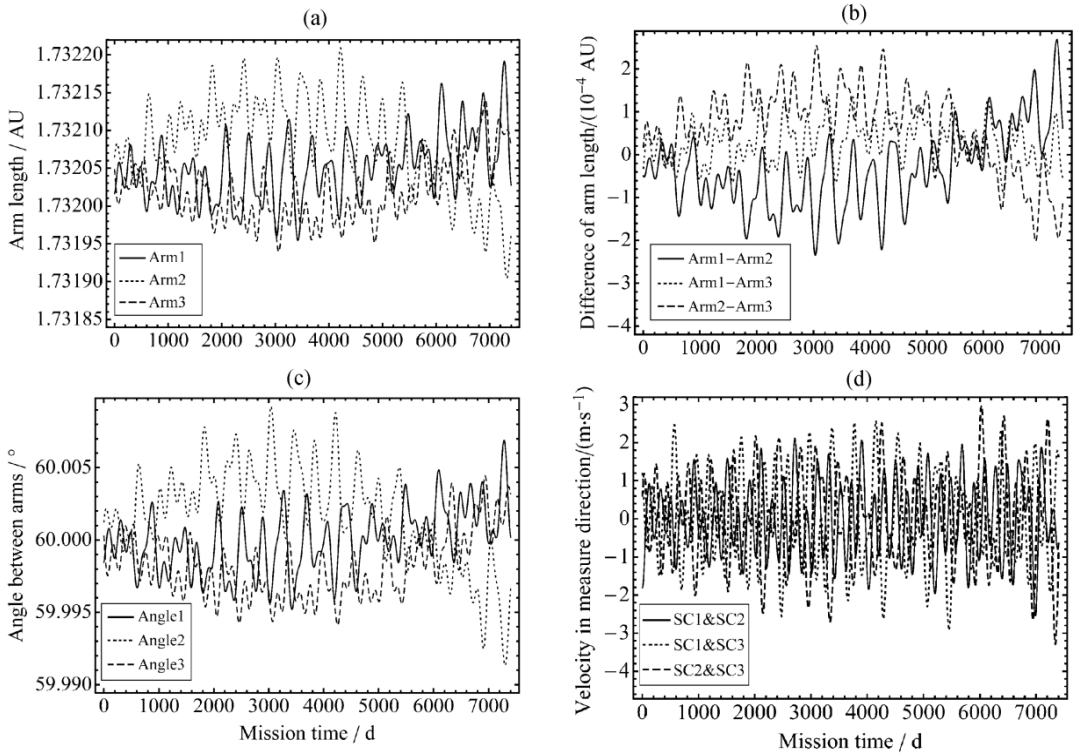

Fig. 4. The variation of (a) arm lengths, (b) difference of arm lengths, (c) angles between arms and (d) velocities in the measure direction in 20 years.

In the general case, after application of these two steps once, the ASTROD-GW requirements would still not be satisfied. Repeated applications of period and eccentricity optimizations are needed to satisfy the requirements. The initial conditions obtained after period optimization, and after all optimizations are listed for comparison in column 4 and 5 of Table 1 in the J2000 equatorial solar-system-barycentric coordinate system. The variation of arm lengths, difference of arm lengths, angles between arms, and velocities in the line-of-sight direction are drawn in Figure 4 for 20 years

The index label of arm is defined by the index of the opposite spacecraft, and the index label of angle is the same as the index of spacecraft.

**5. Time-delay interferometry (TDI)**

For laser-interferometric antenna for space detection of GWs, the arm lengths vary according to orbit dynamics. In order to attain the requisite sensitivity, laser frequency noise must be suppressed below the secondary noises such as the optical path noise, acceleration noise etc. For suppressing laser frequency noise, time delay interferometry to match the optical path length of different beams is needed. The better match of the optical path lengths are, the better cancellation of the laser frequency noise and the easier to achieve the requisite sensitivity. In case of exact match, the laser frequency noise is fully cancelled, as in the original Michelson interferometer.

In the study of ASTROD mission concept, time-delay interferometry was first used.[21,22,42] In the deep-space interferometry, long distance is invariably involved. Due to long distance, laser light is attenuated to great extent at the receiving spacecraft. To transfer the laser light back or to another spacecraft, amplification is needed. The procedure is to phase lock the local laser to the incoming weak laser light and to transmit the local laser back to another spacecraft. We have demonstrated in the laboratory the phase locking of a local oscillator with 2 pW laser light.[34,35] Dick et al.[36] have demonstrated phase locking to 40 fW incoming weak laser light. The power requirement feasibility for ASTROD-GW is met with these developments. In the 1990s, we used the following two time-delay interferometry configuations during the study of ASTROD interferometry and obtain numerically the path length differences using Newtonian dynamics: [21,22,42]

(i)  Unequal arm Michelson TDI Configuration:
   Path 1: S/C1 → S/C2 → S/C1 → S/C3 → S/C1
   Path 2: S/C1 → S/C3 → S/C1 → S/C2 → S/C1

(ii) Sagnac TDI configuration:
   Path 1: S/C1 → S/C2 → S/C3 → S/C1
   Path 2: S/C1 → S/C3 → S/C2 → S/C1

Here we do the same thing for ASTROD-GW. For the numerical evaluation, we take a common receiving time epoch for both beams, the results would be the very close numerically if we take the same starting time epoch and calculate the path differences. The results of this calculation are shown in Fig. 5. We refer the path S/C1 → S/C2 → S/C1 as *a* and the path S/C1 → S/C3 → S/C1 as *b*. Hence the difference of Path 1 minus Path 2 for the unequal-arm Michelson can be denoted as *ab-ba* ≡ [*a*, *b*].

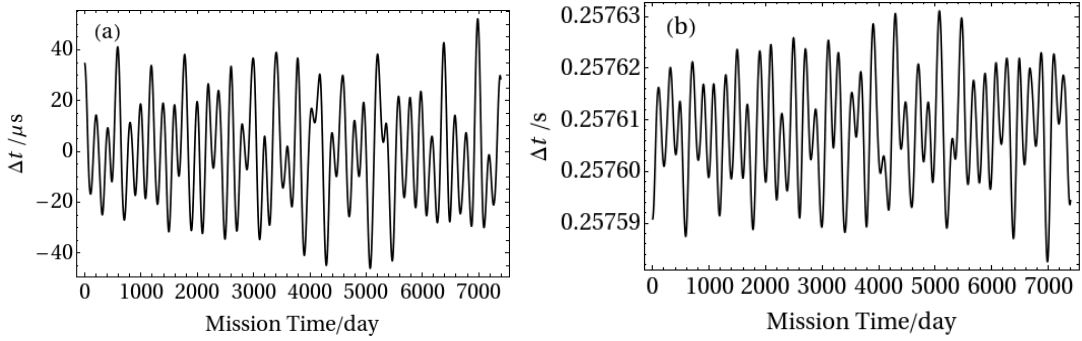

Fig. 5. The path length difference of two optical paths for (a) n = 1 unequal-arm Michelson TDI configuration, i.e., [*a*, *b*] and (b) Sagnac configuration.

Time delay interferometry has been worked out for LISA much more thoroughly since 1999.[23, 24] First-generation and second-generation TDIs are proposed. In the first generation TDIs, static situations are considered, while in the second generation TDIs, motions are compensated to certain degrees. The two configurations considered above and results shown in Fig. 5 are first generation TDI configurations in the sense of Tinto *et al.*, although the calculations are using dynamics. We shall not review more about these historical developments here, but refer the readers to the excellent review of Tinto and Dhurandhar[24] for comprehensive treatment.

In the following section, we will work out the path length differences of second-generation TDIs numerically for the two-arm case of ASTROD-GW.

## 6. Second-generation TDI for ASTROD-GW

The second-generation TDIs obtained by Dhurandhar *et al*[26] are listed in degree-lexicographic order as following:

(i) n=1, [*ab*, *ba*] = *abba* – *baab*

(ii) n=2, [$a^2b^2$, $b^2a^2$]; [*abab*, *baba*]; [$ab^2a$, $ba^2b$]

(iii) n=3, [$a^3b^3$, $b^3a^3$], [$a^2bab^2$, $b^2aba^2$], [$a^2b^2ab$, $b^2a^2ba$], [$a^2b^3a$, $b^2a^3b$],

[$aba^2b^2$, $bab^2a^2$], [*ababab*, *bababa*], [$abab^2a$, $baba^2b$], [$ab^2a^2b$, $ba^2b^2a$],

[$ab^2aba$, $ba^2bab$], [$ab^3a^2$, $ba^3b^2$], lexicographic (binary) order

For the numerical evaluation, we take a common receiving time epoch for both beams and calculate the path differences. Fig. 5, Fig. 6 and Fig. 7 show the numerical results for the n = 1, n = 2 and n = 3 TDI configurations respectively.

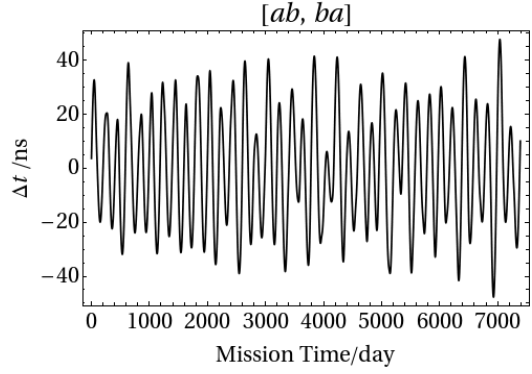

Fig. 5. The path length difference of two optical paths of n = 1 TDI configuration.

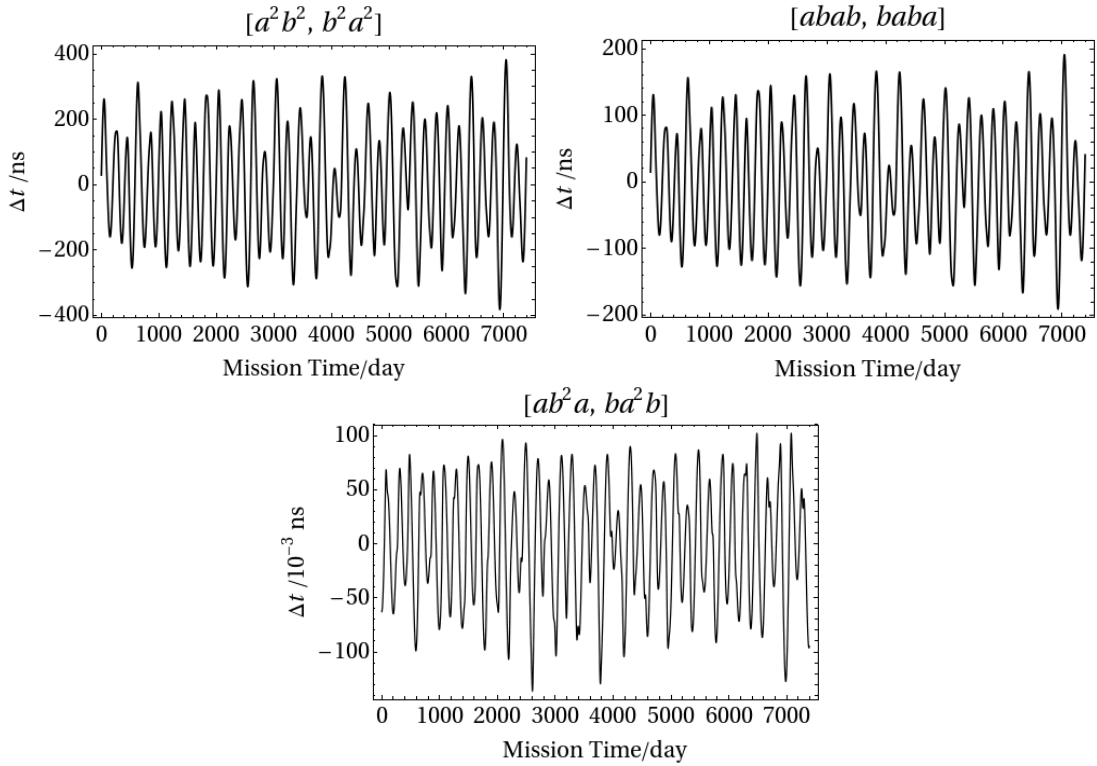

Fig. 6. The path length difference of two optical paths of three n = 2 TDI configurations.

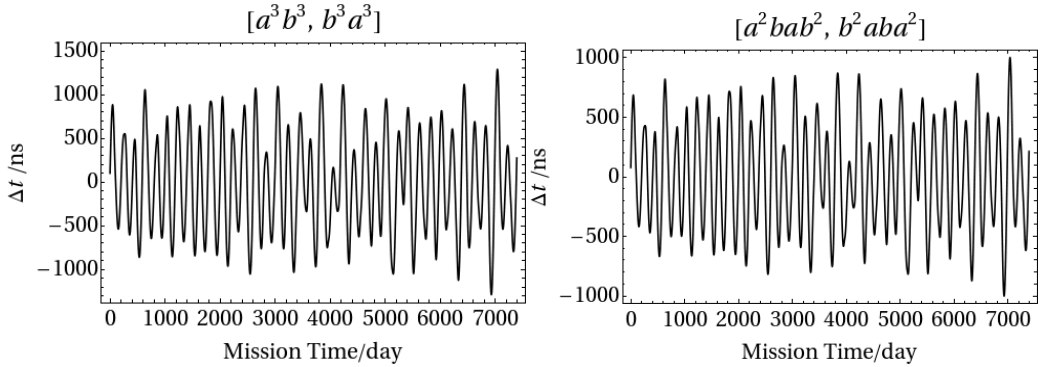

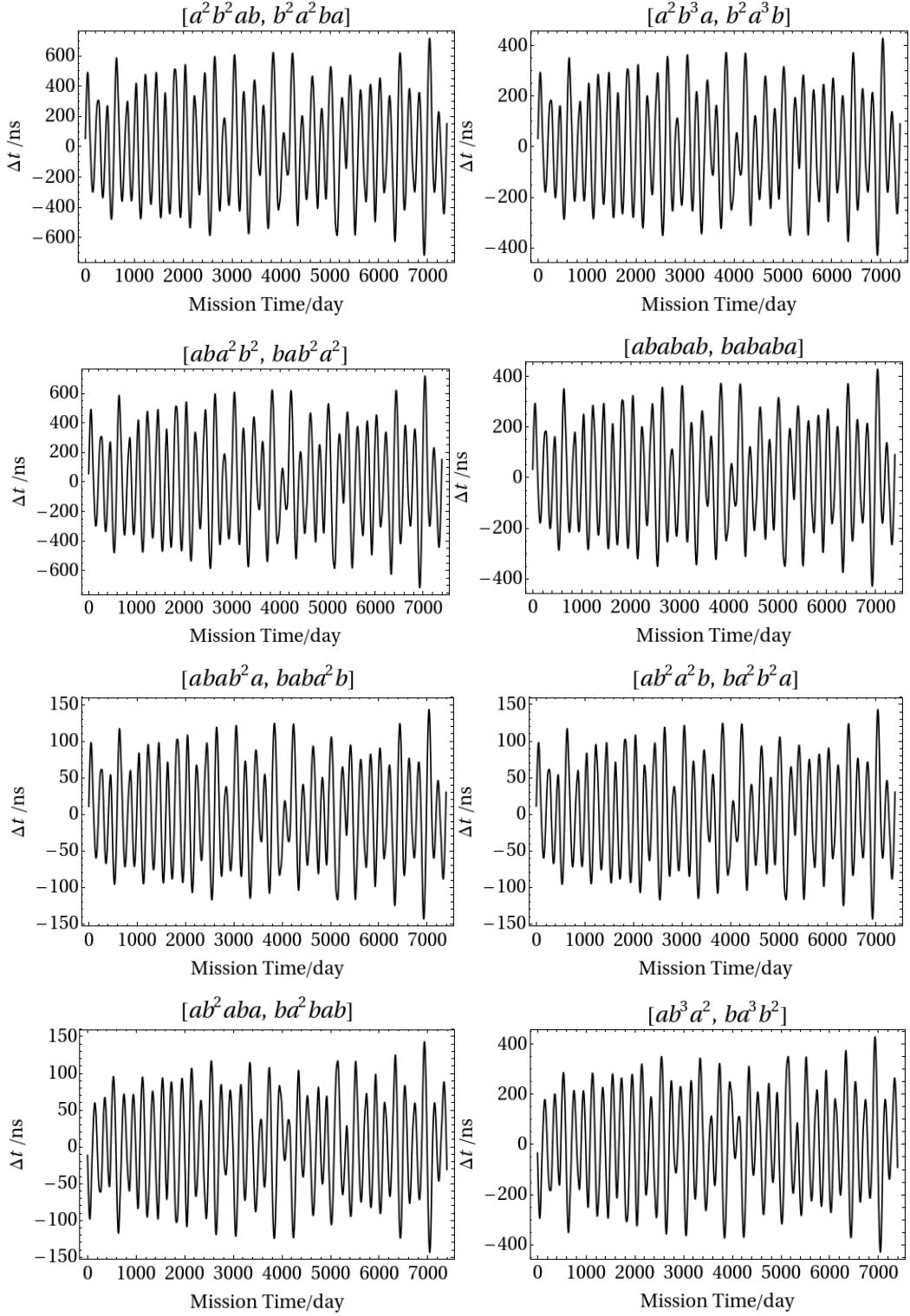

Fig. 7. The path length difference of two optical paths of ten n = 3 TDI configurations.

## 8. Discussions

In the ASTROD-GW mission for detecting GWs in the frequency range 100 nHz-100 mHz, the 3 spacecraft range interferometrically with one another with arm length about 260 million kilometers. After optimization of orbits, changes of arm length can be less than 0.0003 AU and the relative Doppler velocities can be less than 3m/s for 20 years of mission. In order to attain the requisite sensitivity for ASTROD-GW, laser frequency noise must be suppressed below the secondary noises such as the optical path noise, acceleration noise etc. For suppressing laser frequency noise, we use time delay interferometry to match the time (optical path length). To conform to the ASTROD-GW planning, we work out a set of 20-year optimized mission orbits of ASTROD-GW spacecraft starting at June 21, 2028, and calculate the residual errors in the first and second generation time delay interferometry. All the second generation time delay interferometry calculated in this paper satisfies the ASTROD-GW requirement of less than 500 m.

In this paper, except for Sagnac TDI configuration, we have only considered one-detector two-arm cases. Other first-generation numerical TDI solutions and some of the second-generation numerical TDI solutions for full three arms are worked out in [43]. Comparisons with TDIs for NGO/eLISA and LISA are discussed in [28].

**Acknowledgements**

We would like to thank the National Natural Science Foundation (Grant Nos 10778710 and 10875171) for supporting this work in part.